\documentclass{sig-alternate}
\usepackage{color}

\usepackage{algorithm}
\usepackage{algorithmic}
\usepackage{pifont}

\usepackage[colorlinks]{hyperref}
\usepackage[colorinlistoftodos]{todonotes}
\begin{document}

% Standard definitions

% New environment definition

% The following data (volume, number and page) are given by the editors prior to publishing your article

% Includes headers with simplified name of the authors and article title

% Title of the article
\title{Is Learning to Rank Worth it? \\ A Statistical Analysis of Learning to Rank Methods}

% List of authors
\numberofauthors{4}

\author{
% 1st. author
\alignauthor
       Guilherme de C. M. Gomes\\
       \affaddr{Dep. of Computer Science}\\
       \affaddr{Fed. Univ. Minas Gerais}\\
       \affaddr{Belo Horizonte, Brazil}\\
       \email{gcm.gomes@dcc.ufmg.br}
% 2nd. author
\alignauthor
       Vitor C. Oliveira\\
       \affaddr{Dep. of Computer Science}\\
       \affaddr{Fed. Univ. Minas Gerais}\\
       \affaddr{Belo Horizonte, Brazil}\\
       \email{vitorco@dcc.ufmg.br}
% 3rd. author
\alignauthor 
       Jussara M. Almeida\\
       \affaddr{Dep. of Computer Science}\\
       \affaddr{Fed. Univ. Minas Gerais}\\
       \affaddr{Belo Horizonte, Brazil}\\
       \email{jussara@dcc.ufmg.br}
% use '\and' if you need 'another row' of author names
% 4th. author
\and \alignauthor 
       Marcos A. Gon\c{c}alves\\
       \affaddr{Dep. of Computer Science}\\
       \affaddr{Fed. Univ. Minas Gerais}\\
       \affaddr{Belo Horizonte, Brazil}\\
       \email{mgoncalv@dcc.ufmg.br}
}

\date{4 February 2013}

% THE ARTICLE BEGINS

\maketitle

% Article abstract - it should be from 100 to 300 words
\begin{abstract} The Learning to Rank (L2R) research field has experienced a fast paced growth over the last few years, with a wide variety of benchmark datasets and baselines available for experimentation. We here investigate the main assumption behind this field, which is that, the use of sophisticated L2R algorithms and models, produce significant gains over more traditional and simple information retrieval approaches. Our experimental results surprisingly indicate that many L2R algorithms, when put up against the best individual features of each dataset, may not produce statistically significant differences, even if the absolute gains may seem large. We also find that most of the reported baselines are statistically tied, with no clear winner. \end{abstract}

% ACM Computing Classification System categories

% Article keywords
\section{Introduction}

Over the last few years, Learning to Rank (L2R) has become a very popular research topic, based on the general and well-accepted assumption that it produces a much better performance than traditional ranking methods, such as BM25 \cite{[1]} or Language Based Models \cite{[2]}, in information retrieval tasks. Indeed, several new L2R methods \cite{[3]} and benchmark datasets, including large ones such as the LETOR repository \cite{[7]}, have been developed and made available to the community, in recent years.

However, the development and efficient employment of such methods are not free of costs. Being based on supervised learning, they require labeled datasets in order to properly learn the ranking functions. Moreover, these datasets should be large and heterogeneous enough to be capable of representing the domains upon which they will be applied. Due to such strict requirements, constructing such datasets is not a trivial task. In fact, it is very costly. After building the required data, an usually very computationally demanding learning phase has to be applied to learn the ranking functions, which may also require an expensive parameter tuning for optimal performance. Finally, the use of such functions, in production mode in real search engines for example, is usually a two-stage process, in which traditional methods are first applied and, in a subsequent step, the more expensive learned function is used to re-rank the top results generated by the first step \cite{[4]}. This implies in an additional overhead to produce query answers.

Given all these issues, as well as the continuous advance and interest in the area, we here take a step back and reevaluate the main assumption upon which Learning to Rank built its foundations, which is that  the use of sophisticated L2R algorithms and models  produce significant gains over more traditional and simple information retrieval approaches. We also investigate whether there is one (or more) algorithm, out of various L2R techniques  that have been proposed in the literature, that deliver superior effectiveness in most situations (e.g., different collections, different tasks, etc). In order to do so, we analyze the results of 13 methods, here referred to as baselines, over 6 large datasets of the LETOR 3.0 benchmark \cite{[7]}, as well as 6 baselines over 2 even larger datasets of the LETOR 4.0 benchmark \cite{[7]}, when put up against simple isolated feature rankers, using statistically significance tests. All the datasets and baseline results (but one) are available at the benchmark's web page \cite{[7]}. The only new method  we tested that is not in the LETOR benchmark is a new implementation of a Random Forest  ranker \cite{[12]}, which, as we shall see,   produced some  good results in some of our  experiments. Our goal is to verify whether the effectiveness of these methods is better than that produced with the best feature of each dataset when used in isolation, as given by some measure of ranking quality (e.g., Mean Average Precision). We also contrast the performance of each method against each other using the same statistical methodology and datasets. To our knowledge no previous work has performed such detailed comparison with a rigorous statistical analysis.

Our experimental results show that: (1) in most datasets, the best single feature, ranked by Mean Average Precision (MAP) or Normalized Discounted Cumulative Gain cut at the top 10 element (NDCG@10) \cite{[14]} \cite{[15]}, produces results that are statistically tied to most of the reported baselines; (2) the absolute differences in effectiveness provided by the L2R algorithms, when compared to single feature rankers, may be large, but, in most cases, are not statistically significant; and (3) almost all the baselines have very similar performances, making it unlikely that there is an overall best L2R method. Therefore a clear advantage of L2R solutions may not be confirmed in all situations, mainly considering the costs involved.

In comparison to its preliminary version \cite{[16]}, this paper brings new analyses of all benchmark algorithms in all datasets using a different evaluation metric, namely Normalized Discounted Cumulative Gain (NDCG) \cite{[14]} \cite{[15]}, and includes a new L2R algorithm that is not present in the LETOR benchmark, namely Random Forests, which  demonstrated potentially good performance in several of our experiments.

The remainder of this paper is organized as follows: Section~\ref{sec:Related_Work} describes work related to this paper; Section~\ref{sec:Setup} details our experiments and analyses; Section~\ref{sec:Results} reports our results; Section~\ref{sec:Conclusions} concludes the paper and describes our future work.

\section{Related Work}
\label{sec:Related_Work}
Despite the great interest in Learning to Rank in recent years, most of the related work focuses on proposing new algorithms for ranking or novel applications of existing ones. After the publication of the LETOR dataset \cite{[8]}, very few studies were made concerning the effects of these public datasets on the task of learning to rank.

In \cite{[9]}, the authors observed that the ways in which documents were selected for each topic of the LETOR benchmark presented on \cite{[8]} show that the selection has (for each of the three corpora) a particular bias or skewness. This observation has some unexpected effects that may considerably influence any learning-to-rank exercise conducted on these datasets. However, most of these problems were explained and corrected by the benchmark's authors in \cite{[10]}. In \cite{[11]}, a comparison of 7 learning to rank algorithms is made on the LETOR 3.0 benchmark. Each algorithm is compared with each other in terms of Mean Average Precision (MAP) and Normalized Discounted Cumulative Gain (NDCG). In comparison with this previous study, we here compare 13 learning to rank algorithms against not only each other but also against using the best single feature in various datasets of the LETOR 3.0 and 4.0 benchmarks. Moreover, unlike most previous work,  we use statistical tests to support our analyses and conclusions.

An interesting study is reported in \cite{[13]} whose goal is not to improve L2R algorithms but instead to reduce the cost of producing labeled datasets for the learning process. An unexpected finding was that the reduced training sets produced better results than using the whole datasets with some of the tested rankers in the LETOR datasets. This result points towards an interesting line of research on removing noise and redundancy in L2R datasets.

However, none of these previous efforts effectively evaluated the real gains of learning algorithms over traditional methods, like BM25 or Language Models, expressed as features of the dataset. Moreover, to our knowledge,  the first effort  towards  a statistical comparison of learning to rank algorithms and an evaluation of their differences against the best single features was our previous work \cite{[16]}. This   work extends this preliminary analysis by  performing a more thorough evaluation, considering different evaluation metrics and including one more recently proposed L2R algorithm, and  thus reaching more solid conclusions. 

\section{Experimental Setup}
	\label{sec:Setup}
	For our experimentations, we employed 8 datasets from the LETOR benchmark \cite{[7]}. Namely, we used the HP2003, HP2004, NP2003, NP2004, TD2003, and TD2004 datasets, from LETOR 3.0, based on the Gov web page collection. The first four collections are more related to navigational tasks, in which a single unique page is the sole best answer for a query, while the latter two are related to more traditional informational queries. We also use the larger MQ2007, MQ2008 datasets from LETOR 4.0, based on the Gov2 collection, which are also related to informational tasks. All datasets are divided into 5 folds, with the goal of performing a 5-fold cross-validation, that is, 3 folds are used for training, one (validation) for parameter tuning, and the remaining one for testing. In the following, we start by describing in more detail each group of datasets in Section~\ref{sec:Collections} and then further describe our experimental methodology in Section~\ref{sec:Choice}.

	\subsection{Collections and Baselines}
		\label{sec:Collections}
		%\subsubsection{LETOR 3.0}
		Each of the many datasets encapsulated by the LETOR 3.0 benchmark is composed of feature vectors for query-document pairs, along with a corresponding relevance judgment indicating whether the document is relevant or not for the query. There are 64 features per pair, which correspond to various pieces of information commonly used by traditional approaches (such as PageRank) or the result of directly applying simpler methods, such as TF*IDF, BM25 and language models, for estimating the document's relevance to the query. Considering all 6 datasets contained within this version of the benchmark, we find 575 queries and over 580.000 labeled documents.

		Also available on the benchmark's web page, we find 12 different baselines, namely: FRank, ListNet, AdaRank-MAP, AdaRank-NDCG, RankBoost, RankSVM, RankSVM-Struct, RankSVM-Primal, Regression, Regression+L2reg, Smooth Rank and SVMMAP. Aside from FRank and RankBoost, all algorithms use linear ranking functions.

		%\subsubsection{LETOR 4.0}
		Similar to the previous benchmark, LETOR 4.0 uses the same structure for its datasets, but with 46 features, instead of 64, per vector. The largest TREC datasets available, from the Million Query Tracks of 2007 and 2008, named as MQ2007 and MQ2008, are based on the Gov2 webpage collection. Both datasets sum over 2.500 queries and roughly 420.000 documents. Unlike in LETOR 3.0, we here find only 5 reported baseline algorithms: AdaRank-MAP, AdaRank-NDCG, ListNet, RankBoost and RankSVM-Struct. Any other information pertaining these baselines and datasets, as well as the datasets themselves, are available at the LETOR website \cite{[7]}.
		
		Aside from the reported LETOR benchmark algorithms, we have also performed experiments of our own using Random Forests, as implemented in \cite{[12]}, and included its results in our analyses, making it the $13^{th}$ and $6^{th}$ baseline method for LETOR~3.0 and 4.0, respectively. As we shall see,  its performance is very similar to the other methods, even though the Random Forest approach is quite different from the other baselines, making it an interesting comparison candidate. As for the parameterization, we used only the validation sets, as  done for the LETOR baselines. 
		
		%That is, we did not perform any algorithm-specific tuning, such as setting the number of used bags or the number of applied trees in the learning phase.

	\subsection{Choice of the Best Feature}
		\label{sec:Choice}
	For comparison purposes, we performed the following feature selection procedure. For each fold, the values of each feature of each query-document pairs of the test set are extracted and used as a ranking score for its respective pair, thus obtaining a number of ranked lists equal to the number of features used to describe each document. Afterwards, we use the evaluation tool provided by the benchmark to calculate traditional information retrieval metrics, such as Mean Average Precision (MAP) and average Normalizing Discounted Cumulative Gain (NDCG) \cite{[14]} \cite{[15]}, of the previously produced  ranked lists. We then select the feature responsible for the best ranked list, in terms of generated MAP and average NDCG@10, to use as our isolated ranking feature.

	We here compare both MAP and average NDCG@10 results with those obtained by the L2R baselines in the same test sets, using statistical significance tests with a 95\% confidence level, aiming at quantifying the differences and verifying whether they are statistically significant. Specifically, to support our analyses and conclusions, we perform a pairwise comparison of all methods, applying paired difference tests \cite{[5]} for each pair, to verify whether they are statistically different.

\section{Results}
	\label{sec:Results}
	We start by showing the best results obtained with a single feature for each considered dataset and comparing these results with the analyzed L2R algorithms.

	In order to perform a more thorough investigation on the effectiveness of the  Learning to Rank algorithms, we first evaluate results of the single best feature ranking method when compared to the L2R baselines using MAP (Section~\ref{sec:FRR_MAP}) and NDCG@10 (Section~\ref{sec:FRR_NDCG@10}) as our feature selection metric .
	Next, we compare all baselines using both MAP (Section~\ref{sec:COMP_MAP}) and NDCG@10 (Section~\ref{sec:COMP_NDCG@10}).

	\subsection{Feature Ranking Results - MAP}
		\label{sec:FRR_MAP}
		As a result of our feature selection procedure, a single attribute was selected as the best one in each dataset. Some features were selected for more than one dataset. Table~\ref{tab:F_MAP} shows the MAP scores generated by the evaluation tools for each single feature, as well as the feature name and its position in the document vectors. In particular, it is interesting to notice the lower MAP values in the TD collections, which are known to be difficult informational datasets.

	\begin{table}[!htb]
			\begin{center}
				\centering \scriptsize
				\caption{Average MAP scores for the best ranked features of the different datasets, LETOR 3.0}
				\label{tab:F_MAP}
				\begin{tabular}{|l|c|p{4cm}|}
					\hline
					Dataset & MAP Score & Best Feature \\
					\hline
					HP2003 & 0.7031 & Hyperlink base feature propagation: weighted in-link (46)\\
					\hline
					HP2004 & 0.6171 & Hyperlink base feature propagation: weighted in-link (46)\\
					\hline
					NP2003 & 0.5784 & IDF of the URL (9)\\
					\hline
					NP2004 & 0.5202 & Sitemap based score propagation (42)\\
					\hline
					TD2003 & 0.1973 & Hyperlink base feature propagation: weighted in-link (46)\\
					\hline
					TD2004 & 0.1844 & Sitemap based score propagation (42)\\
					\hline
					MQ2007 & 0.4534 & LMIR.DIR of whole document (39)\\
					\hline
					MQ2008 & 0.4712 & LMIR.DIR of whole document (39)\\
					\hline
				\end{tabular}
			\end{center}
		\end{table}

		Table~\ref{tab:F_COMP_L3_MAP} shows the relative MAP difference between the best ranking feature of each dataset and the L2R baselines reported for LETOR 3.0 as well as the Random Forest ranker. Next to the dataset's name, in parenthesis, we find the best feature's identifier. When the performance of the L2R algorithm is statistically better than the single feature ranking, a (+) sign is provided after the MAP difference. When it is statistically equal, (=) is included, whereas a (-) sign indicates that the L2R method is statistically inferior to the single feature. In other words, a positive sign  indicates that the L2R approach has a MAP score significantly higher, with 95\% confidence,  than the single feature, while a negative sign indicates a statistically significant lower   score.

	\begin{table*}[!htb]
			\begin{center}
				\centering \scriptsize
				\caption{Relative MAP comparison between feature ranking and L2R algorithms, LETOR 3.0}
				\label{tab:F_COMP_L3_MAP}
				\centerline{\begin{tabular}{|p{1.7cm}|c|c|c|c|c|c|}
					\hline
					\textbf{Dataset} &\textbf{HP2003 (46)} &\textbf{HP2004 (46)} &\textbf{NP2003 (9)} &\textbf{NP2004 (42)} &\textbf{TD2003 (42)} &\textbf{TD2004 (42)} \\
					\hline
					\textbf{AdaRank-MAP} & 9.65\% (+) & 16.97\% (+) & 17.28\% (+) & 19.56\% (=) & 15.70\% (=) & 18.72\% (=) \\
					\hline
					\textbf{AdaRank-NDCG} & 6.38\% (=) & 12.02\% (=) & 15.46\% (+) & 20.51\% (=) & 20.00\% (=) & 5.00\% (=) \\
					\hline
					\textbf{FRank} & 0.90\% (=) & 10.45\% (+) & 14.8\% (+) & 15.49\% (=) & 2.93\% (=) & 29.56\% (=) \\
					\hline
					\textbf{ListNet} & 8.9\% (=) & 11.78\% (=) & 19.22\% (+) & 29.17\% (+) & 39.52\% (+) & 21.04\% (+) \\
					\hline
					\textbf{RankBoost} & 4.25\% (+) & 1.29\% (=) & 22.31\% (+) & 8.42\% (=) & 15.23\% (=) & 41.77\% (+) \\
					\hline
					\textbf{RankSVM} & 5.35\% (+) & 8.15\% (=) & 20.28\% (+) & 26.64\% (+) & 33.17\% (=) & 21.36\% (=) \\
					\hline
					\textbf{RankSVM-Primal} & 8.72\% (+) & 8.75\% (=) & 19.00\% (+) & 29.85\% (+) & 34.44\% (=) & 11.81\% (=) \\
					\hline
					\textbf{RankSVM-Struct} & 8.45\% (+) & 9.93\% (=) & 17.36\% (+) & 30.17\% (+) & 37.48\% (=) & 19.1\% (+) \\
					\hline
					\textbf{Regression} & -29.35\% (-) & -14.84\% (-) & -2.42\% (=) & -1.16\% (=) & 22.08\% (=) & 12.72\% (=) \\
					\hline
					\textbf{Regression-L2reg} & 6.47\% (=) & 2.09\% (=) & 17.98\% (+) & 31.98\% (+) & 23.33\% (+) & 8.023\% (=) \\
					\hline
					\textbf{SmoothRank} & 5.54\% (=) & 16.28\% (=) & 18.76\% (+) & 27.26\% (+) & 23.93\% (+) & 11.14\% (=) \\
					\hline
					\textbf{SVMMAP} & 8.56\% (=) & 16.24\% (+) & 20.31\% (+) & 29.95\% (+) & 36.61\% (+) & 26.14\% (+) \\
					\hline
					\textbf{Random Forest}& 9.41\% (+) & 2.18\% (=) & 21.39\% (+) & 15.32\% (=) & 38.35\% (+) & 38.3\% (+) \\
					\hline
				\end{tabular}}
			\end{center}
		\end{table*}

		By looking at Table~\ref{tab:F_COMP_L3_MAP}, we see that, aside from the NP datasets, only  half of the L2R algorithms are statistically superior to the isolated feature ranking in each dataset. In other words, in several cases the absolute differences in performance may not be statistically sound. In fact, despite some large {\it average} gains, there are a lot of statistical ties and even performance losses of the L2R algorithms, in comparison with the isolated feature rankings. For instance, in the HP2004 dataset, a difference (on average) of 16.28\% of the SmoothRank algorithm over the best single feature is indeed not significant:  both methods are tied with 95\% confidence. This is very surprising as we expected that all or at least most of the algorithms would be able to effectively combine the features to deliver a better performance. However, the variability of the results is so large that relying only on average MAP to determine the best method is not enough.
In contrast, a 10.45\% gain of the FRank method over the best single feature in the same dataset is significant.  Since the number of replications and confidence level are  fixed (i.e., 5 and 95\%, respectively), the lower variability inherent to FRank results allows us to conclude towards a significant difference. 

For the NP2003 and NP2004 datasets, most L2R algorithms (92.3\% and 53.8\% of the considered methods, respectively) are indeed statistically superior to the best single feature. However, it is interesting to note that, even in these datasets, some (apparently) large relative differences (e.g., 19.56\%) are in fact not statistically significant with 95\% confidence. It is also worth mentioning the large and significant losses (up to 29.35\%) of the Regression method over the best feature in the two HP datasets. This may me due to the high correlations among several features in this dataset, which may be detrimental to this particular method as it relies on linear regression \cite{[5]}.

		Regarding the Random Forest algorithm, Table \ref{tab:F_COMP_L3_MAP} shows that it outperforms the best single feature approach, with statistically significant gains, in 4 out of the 6 analyzed datasets, being statistically tied in the other two datasets.  Some of the gains are quite large (over 38\%), such as the ones in the TD2003 and TD2004 datasets. 

		Analogous to Table~\ref{tab:F_COMP_L3_MAP}, Table~\ref{tab:F_COMP_L4_MAP} presents results relative to LETOR 4.0. Similarly to the results found in LETOR 3.0, we here see cases where not only the relative differences are not statistically significant (with 95\% confidence), such as in MQ2008, but also   some cases where the gains are only marginal (e.g., 0.96\% for AdaRank-MAP in the MQ2007 dataset). In fact, it is very surprising that in MQ2008, no method is able to outperform the best feature in isolation.

                We note that, for both LETOR~3.0 and 4.0, the Random Forest ranker tends to have a similar behavior as the best baseline, in terms of the gains over the best feature approach, in most datasets but HP2004, MQ2007, and, to a lesser extent, NP2004.  In other words, in all other datasets,  the relative difference between the Random Forest algorithm and the best feature ranker tends to be (close to) the largest  across all considered algorithms, even though no fine parameterization was performed, meaning that this algorithm has a lot of potential.

%                which raises questions as to the potential of the optimal model for each dataset.

	%	Also of interest, both in LETOR~3.0 and 4.0, is the Random Forest behavior. Aside from the HP2004 and MQ2007 datasets, it tends to perform quite similarly to the best baseline (i.e., the one with the highest absolute difference to the single feature), even though no fine parametrization has been performed, as mentioned in Section~\ref{sec:Collections}, raising questions about the potential of the optimal model for each dataset.
		
		%For instance, the high gains observed in the TDs and NPs is also found in the algorithm's results, whereas the low values in MQ2008 and appear, even with the lack of fine parameter tunning mentioned in \ref{sec:Setup}, leading to speculations on the optimal Random Forest model for each dataset and its potential results.

	\begin{table}[!htb]
			\begin{center}
				\centering \scriptsize
				\caption{Relative MAP comparison between feature ranking and L2R algorithms, LETOR 4.0}
				\label{tab:F_COMP_L4_MAP}
				\centerline{\begin{tabular}{|c|c|c|}
					\hline
					\textbf{Dataset} & \textbf{MQ2007 (39)} & \textbf{MQ2008 (39)}\\
					\hline
					\textbf{AdaRank-MAP} & 0.96\% (+) & 1.11\% (=)\\
					\hline
					\textbf{AdaRank-NDCG} & 1.51\% (+) & 2.38\% (=)\\
					\hline
					\textbf{ListNet} & 2.61\% (+) & 1.34\% (=)\\
					\hline
					\textbf{RankBoost} & 2.83\% (+) & 1.34\% (=)\\
					\hline
					\textbf{RankSVM-Struct} & 2.45\% (+) & -0.34\% (=)\\
					\hline
					\textbf{Random Forest} & 1.07\% (=) & 0.47\% (=)\\
					\hline
				\end{tabular}}
			\end{center}
		\end{table}

		These results lead to interesting conclusions pertaining the effective gains of L2R and its aggregated costs. While the performed process of choosing the best features is not cost free, it is much cheaper than the complex machine learning algorithms. Indeed, we may not need to investigate all possible features. A smaller set of candidates could be used based on results reported in the literature.

	\subsection{Feature Ranking Results - NDCG@10}
		\label{sec:FRR_NDCG@10}
		Like Table~\ref{tab:F_MAP}, Table~\ref{tab:F_NDCG@10} shows average NDCG@10 scores of the selected best feature for each dataset. Interestingly, 6 out of 8 selections match the ones made using MAP as our attribute selection metric.   Moreover, 4 out of the 6 datasets in LETOR 3.0 have feature 46 (Hyperlink base feature propagation: weighted in-link) as the top ranking feature. Features that were chosen using both MAP and NDCG@10 are shown in bold in Table \ref{tab:F_NDCG@10}.
	
	\begin{table}[!htb]
		\begin{center}
			\centering \scriptsize
			\caption{Average NDCG@10 scores for the best ranked features of the different datasets, LETOR 4.0}
			\label{tab:F_NDCG@10}
			\begin{tabular}{|l|c|p{4cm}|}
				\hline
				Dataset & NDCG@10 Score & Best Feature \\
				\hline
				HP2003 & 0.7910 & \textbf{Hyperlink base feature propagation: weighted in-link (46)}\\
				\hline
				HP2004 & 0.7511 & \textbf{Hyperlink base feature propagation: weighted in-link (46)}\\
				\hline
				NP2003 & 0.6907 & LMIR.ABS of whole document (30)\\
				\hline
				NP2004 & 0.6125 & \textbf{Sitemap based score propagation (42)}\\
				\hline
				TD2003 & 0.2637 & \textbf{Hyperlink base feature propagation: weighted in-link (46)}\\
				\hline
				TD2004 & 0.2748 & Hyperlink base feature propagation: weighted in-link (46) \\
				\hline
				MQ2007 & 0.4221 & \textbf{LMIR.DIR of whole document (39)}\\
				\hline
				MQ2008 & 0.2230 & \textbf{LMIR.DIR of whole document (39)}\\
				\hline
			\end{tabular}
		\end{center}
	\end{table}

		Table~\ref{tab:F_COMP_L3_NDCG@10} shows the relative difference in average NDCG@10 and statistical significance (i.e., (+), (-) or (=)) between each L2R method  and the single feature procedure in each dataset of LETOR 3.0. All values and statistical results were computed as in Table~\ref{tab:F_COMP_L3_MAP}. Algorithms that showed a similar significance both in MAP and NDCG@10 results are displayed in bold.

	\begin{table*}[!htb]
		\begin{center}
			\centering \scriptsize
			\caption{Relative average NDCG@10 comparison between feature ranking and L2R algorithms, LETOR 3.0}
			\label{tab:F_COMP_L3_NDCG@10}
			\centerline{\begin{tabular}{|p{1.7cm}|c|c|c|c|c|c|}
			\hline
			\textbf{Dataset} &\textbf{HP2003 (46)} &\textbf{HP2004 (46)} &\textbf{NP2003 (30)} &\textbf{NP2004 (42)} &\textbf{TD2003 (46)} &\textbf{TD2004 (46)} \\
			\hline
			\textbf{AdaRank-MAP} & \textbf{6.00\% (+)} & \textbf{10.88\% (+)} & \textbf{10.63\% (+)} & \textbf{22.40\% (=)} & \textbf{16.37\% (=)} & \textbf{19.54\% (=)} \\
			\hline
			\textbf{AdaRank-NDCG} & \textbf{1.89\% (=)} & \textbf{7.27\% (=)} & \textbf{11.09\% (+)} & 20.55\% (+) & \textbf{15.13\% (=)} & \textbf{15.10\% (=)} \\
			\hline
			\textbf{FRank} & 0.75\% (+) & 1.39\% (=) & \textbf{12.41\% (+)} & \textbf{19.12\% (=)} & \textbf{2.01\% (=)} & \textbf{21.25\% (=)} \\
			\hline
			\textbf{ListNet} & 5.84\% (+) & \textbf{4.45\% (=)} & \textbf{16.09\% (+)} & \textbf{32.70\% (+)} & \textbf{32.10\% (+)} & 15.56\% (=) \\
			\hline
			\textbf{RankBoost} & 3.30\% (=) & \textbf{-1.10\% (=)} & \textbf{16.82\% (+)} & \textbf{12.88\% (=)} & \textbf{18.39\% (=)} & \textbf{27.53\% (+)} \\
			\hline
			\textbf{RankSVM} & 2.12\% (=) & \textbf{2.35\% (=)} & \textbf{15.88\% (+)} & \textbf{31.62\% (+)} & 31.25\% (+) & \textbf{12.03\% (=)} \\
			\hline
			\textbf{RankSVM-Primal} & \textbf{3.41\% (+)} & \textbf{2.78\% (=)} & \textbf{14.30\% (+)} & \textbf{29.80\% (+)} & 35.42\% (+) & \textbf{6.01\% (=)} \\
			\hline
			\textbf{RankSVM-Struct} & 3.18\% (=) & \textbf{2.07\% (=)} & \textbf{15.18\% (+)} & \textbf{30.23\% (+)} & 31.47\% (+) & 12.47\% (=) \\
			\hline
			\textbf{Regression} & \textbf{-24.86\% (-)} & \textbf{-13.88\% (-)} & \textbf{-3.58\% (=)} & \textbf{6.70\% (=)} & \textbf{23.74\% (=)} & \textbf{10.3\% (=)} \\
			\hline
			\textbf{Regression-L2reg} & \textbf{3.86\% (=)} & \textbf{-4.30\% (=)} & \textbf{16.19\% (+)} & \textbf{31.27\% (+)} & 25.03\% (=) & \textbf{3.06\% (=)} \\
			\hline
			\textbf{SmoothRank} & \textbf{1.06\% (=)} & \textbf{7.34\% (=)} & \textbf{15.46\% (+)} & \textbf{31.89\% (+)} & \textbf{24.47\% (+)} & \textbf{5.78\% (=)} \\
			\hline
			\textbf{SVMMAP} & 5.24\% (+) & 9.45\% (=) & \textbf{15.63\% (+)} & \textbf{31.83\% (+)} & 27.67\% (=) & \textbf{21.67\% (+)} \\
			\hline
			\textbf{Random Forest}& \textbf{4.98\% (+)} & \textbf{-4.65\% (=)} & \textbf{15.04\% (+)} & \textbf{15.63\% (=)} & \textbf{34.14\% (+)} & \textbf{27.29\% (+)}\\
			\hline
			\end{tabular}}
		\end{center}
	\end{table*}

	Note that 63 out of the 78 average NDCG@10 results (80.8\%) reported in Table \ref{tab:F_COMP_L3_NDCG@10} have the same statistical behavior as the ones reported for MAP. 
 Indeed, the same overall conclusions regarding the lack of a direct correspondence between large absolute gains and statistically significant differences, drawn based on MAP results, also hold for NDCG@10 results. For example, a relative difference of 3.86\% of Regression-L2Reg over the best single feature ranking on HP2003 is indeed a statistical tie, whereas the marginal 0.75\% average gains of FRank over the best feature approach on the same dataset  - a much smaller relative difference - is a statistical win.  Once again, the diverse variability inherent to each algorithm plays a key role in these conclusions.
Overall, we find that the L2R algorithm and the best single feature ranking are statistically tied in 47.4\% of the cases, whereas in only 51\% of the cases the former is statistically superior to the much simpler and cheaper feature selection and ranking approach.

 %there is a statistical tie between the  single feature ranking and the  L2R algorithm, whereas the L2R algorithm nd a superiority rate of only 51\% of the L2R approaches over a much simpler and cheaper method such as the feature selection and ranking.

%	In accordance to the observations regarding the lack of a direct correspondence between high absolute gains and statistical differences, made in Section~\ref{sec:FRR_MAP}, similar observations can also been made for the NDCG@10 results, including for the Random Forest ranker. 
	
%\textbf{In fact, every NDCG@10 result agrees with that of MAP for the non-standard baseline, showing it's more homogeneous performance in terms of the different evaluated metrics.}

%	When looking at the HP2003, we have that a small difference, 3.86\% of Regression-L2Reg is a tie with the feature ranking method, while FRank's marginal 0.75\%, which is not even 20\% of Regression-L2Reg's difference, is a statistical win. In the TD2003 dataset, for instance, SmoothRank's 27.67\% gain over the single feature ranking approach, is not statistically meaningful, while AdaRank-MAP's 16.37\% (roughly 59\% of SmoothRank's value) is significant. 

Table~\ref{tab:F_COMP_L4_NDCG@10} shows  NDCG@10 results for the LETOR 4.0 benchmark. Once again, these results lead to very similar conclusions as the MAP results,  reported in Table~\ref{tab:F_COMP_L4_MAP}, with the exception  of Random Forest in MQ2007. Unlike observed for MAP, Random Forest is statistically superior to the best single feature in terms of average NDCG@10 in that dataset.
	
	%thus the discussion of its data is the same presented before.

	\begin{table}[!htb]
		\begin{center}
			\centering \scriptsize
			\caption{Relative NDCG@10 comparison between feature ranking and L2R algorithms, LETOR 4.0}
			\label{tab:F_COMP_L4_NDCG@10}
			\centerline{\begin{tabular}{|c|c|c|}
				\hline
				\textbf{Dataset} & \textbf{MQ2007 (39)} & \textbf{MQ2008 (39)}\\
				\hline
				\textbf{AdaRank-MAP} & \textbf{2.71\% (+)} & \textbf{2.45\% (=)}\\
				\hline
				\textbf{AdaRank-NDCG} & \textbf{3.52\% (+)} & \textbf{3.30\% (=)}\\
				\hline
				\textbf{ListNet} & \textbf{5.20\% (+)} & \textbf{3.12\% (=)}\\
				\hline
				\textbf{RankBoost} & \textbf{5.77\% (+)} & \textbf{0.97\% (=)}\\
				\hline
				\textbf{RankSVM-Struct} & \textbf{5.17\% (+)} & \textbf{2.05\% (=)}\\
				\hline
				\textbf{Random Forest}  & 4.16\% (+) & \textbf{2.00\% (=)}\\
				\hline
			\end{tabular}}
		\end{center}
	\end{table}

	\subsection{Baseline Comparisons - MAP}
		\label{sec:COMP_MAP}
		We now turn to our second goal, which is to compare the supervised rankers in the used collections. Tables~\ref{B_COMP_L3_MAP} and \ref{B_COMP_L4_MAP} show  average MAP results obtained for each baseline, jointly with the corresponding 95\% confidence intervals. Best results for each dataset, along with statistical ties according to paired tests with 95\% confidence, are shown in bold. %(the tests were performed by computing 95\% confidence intervals over the differences between the results obtained for each pair of baselines \cite{[5]})

	\begin{table*}[!htb]
		\begin{center}
			\centering \scriptsize
			\caption{Baselines' average MAP and confidence intervals across the different datasets, LETOR 3.0}
			\label{B_COMP_L3_MAP}
			\centerline{\begin{tabular}{|p{1.4cm}|c|c|c|c|c|c|}
				\hline
				\textbf{Dataset} &\textbf{HP2003} &\textbf{HP2004} &\textbf{NP2003} &\textbf{NP2004} &\textbf{TD2003} &\textbf{TD2004} \\
				\hline
				\textbf{AdaRank-MAP} & \textbf{0.771 $\pm$ 0.071} & \textbf{0.722 $\pm$ 0.103} & \textbf{0.678 $\pm$ 0.087} & \textbf{0.622 $\pm$ 0.055} & \textbf{0.228 $\pm$ 0.106} & 0.219 $\pm$ 0.042 \\
				\hline
				\textbf{AdaRank-NDCG} & \textbf{0.748 $\pm$ 0.125} & \textbf{0.691 $\pm$ 0.053} & \textbf{0.668 $\pm$ 0.103} & \textbf{0.627 $\pm$ 0.046} & \textbf{0.237 $\pm$ 0.129} & 0.194 $\pm$ 0.035 \\
				\hline
				\textbf{FRank} & 0.709 $\pm$ 0.077 & \textbf{0.682 $\pm$ 0.112} & \textbf{0.664 $\pm$ 0.082} & \textbf{0.601 $\pm$ 0.112} & \textbf{0.203 $\pm$ 0.089} & \textbf{0.239 $\pm$ 0.042} \\
				\hline
				\textbf{ListNet} & \textbf{0.766 $\pm$ 0.095} & \textbf{0.69 $\pm$ 0.104} & \textbf{0.690 $\pm$ 0.083} & \textbf{0.672 $\pm$ 0.094} & \textbf{0.275 $\pm$ 0.100} & 0.223 $\pm$ 0.006 \\
				\hline
				\textbf{RankBoost} & 0.733 $\pm$ 0.089 & 0.625 $\pm$ 0.015 & \textbf{0.707 $\pm$ 0.040} & 0.564 $\pm$ 0.036 & \textbf{0.227 $\pm$ 0.087} & \textbf{0.261 $\pm$ 0.034} \\
				\hline
				\textbf{RankSVM} & \textbf{0.741 $\pm$ 0.069} & \textbf{0.667 $\pm$ 0.099} & \textbf{0.696 $\pm$ 0.068} & \textbf{0.659 $\pm$ 0.108} & \textbf{0.263 $\pm$ 0.111} & 0.224 $\pm$ 0.035 \\
				\hline
				\textbf{RankSVM-Primal} & \textbf{0.764 $\pm$ 0.087} & \textbf{0.671 $\pm$ 0.096} & \textbf{0.688 $\pm$ 0.078} & \textbf{0.675 $\pm$ 0.122} & \textbf{0.265 $\pm$ 0.109} & 0.206 $\pm$ 0.027 \\
				\hline
				\textbf{RankSVM-Struct} & \textbf{0.763 $\pm$ 0.094} & \textbf{0.678 $\pm$ 0.084} & \textbf{0.679 $\pm$ 0.073} & \textbf{0.677 $\pm$ 0.090} & \textbf{0.271 $\pm$ 0.115} & 0.220 $\pm$ 0.025 \\
				\hline
				\textbf{Regression} & 0.497 $\pm$ 0.042 & 0.526 $\pm$ 0.075 & 0.564 $\pm$ 0.095 & 0.514 $\pm$ 0.064 & \textbf{0.241 $\pm$ 0.083} & 0.208 $\pm$ 0.034 \\
				\hline
				\textbf{Regression-L2reg} & \textbf{0.749 $\pm$ 0.101} & \textbf{0.63 $\pm$ 0.085} & \textbf{0.682 $\pm$ 0.068} & \textbf{0.687 $\pm$ 0.108} & \textbf{0.243 $\pm$ 0.100} & 0.199 $\pm$ 0.024 \\
				\hline
				\textbf{Smooth Rank} & \textbf{0.742 $\pm$ 0.109} & \textbf{0.718 $\pm$ 0.102} & \textbf{0.687 $\pm$ 0.064} & \textbf{0.662 $\pm$ 0.097} & \textbf{0.245 $\pm$ 0.084} & 0.205 $\pm$ 0.024 \\
				\hline
				\textbf{SVMMAP} & \textbf{0.763 $\pm$ 0.096} & \textbf{0.717 $\pm$ 0.077} & \textbf{0.696 $\pm$ 0.060} & \textbf{0.676 $\pm$ 0.071} & \textbf{0.270 $\pm$ 0.093} & 0.233 $\pm$ 0.032 \\
				\hline
				\textbf{Random Forest} & \textbf{0.769 $\pm$ 0.059}& 0.631 $\pm$ 0.123 & \textbf{0.702 $\pm$ 0.042} & \textbf{0.600 $\pm$ 0.072}& \textbf{0.273 $\pm$ 0.101} & \textbf{0.255 $\pm$ 0.045} \\
				\hline
			\end{tabular}}
		\end{center}
	\end{table*}

		In the HP2003 dataset, there is a statistical tie for the best method among 10 out of 13 baselines, i.e., the differences among them are not statistically significant with 95\% confidence. The worst method is Regression, which is inferior to all baselines and even to the best feature in isolation (difference to the best performer, AdaRankMap, of 35\%). Notice however that the second worst method (FRank) is only at most 8\% worse than the best performer. Thus, in general, except for Regression, the differences among all considered baselines are, if significant, relatively small.
		In the HP2004 dataset we also have statistically tied results for 10 baselines. The only baseline that is significantly inferior to the others is (once again) Regression. Unlike for HP2003, Random Forest did not perform  well in this dataset. We here also observe some large differences that are not statistically significant, such as the gap between AdaRank-Map and Regression+L2reg (12.7\%). As discussed before, these results clearly reflect the large variability of the methods across the various folds of the datasets.

		Unlike in HP2004, all methods but Regression are statistically tied in the NP2003 dataset, making it once again impossible to single out a best ranking method. In the NP2004, we have a similar situation with 10 out of 13 methods statistically tied. Surprisingly, RankBoost, which has a good performance in the previous datasets and is one of two best rankers in TD2004 (see below), is tied with Regression as the worst method.

		In the TD2003 dataset, all methods are tied, with no clear winner or loser. An interesting result found here is the very large relative differences (on average) of ListNet over FRank (26.2\%), although they are still statistically tied with 95\% confidence. In the last dataset of LETOR 3.0, TD2004, RankBoost,  FRank, and Random Forest are tied as the best rankers, outperforming the other 10 baselines. In fact, this dataset is the only one with few (three) methods outperforming most of the other algorithms, with some large significant differences in some cases (up to 26\%). In contrast, in the other datasets, almost all considered baselines are tied as the best rankers, with 95\% confidence.

	\begin{table}[!htb]
		\begin{center}
			\centering \scriptsize
			\caption{Relative MAP comparison between feature ranking and L2R algorithms, LETOR 4.0}
			\label{B_COMP_L4_MAP}
			\centerline{\begin{tabular}{|c|c|c|}
				\hline
				\textbf{Dataset} & \textbf{MQ2007} & \textbf{MQ2008}\\
				\hline
				\textbf{AdaRank-MAP} & 0.458 $\pm$ 0.021 & \textbf{0.476 $\pm$ 0.052}\\
				\hline
				\textbf{AdaRank-NDCG} & \textbf{0.460 $\pm$ 0.024} & \textbf{0.482 $\pm$ 0.054}\\
				\hline
				\textbf{ListNet} & \textbf{0.465 $\pm$ 0.020} & \textbf{0.477 $\pm$ 0.053}\\
				\hline
				\textbf{RankBoost} & \textbf{0.466 $\pm$ 0.023} & \textbf{0.477 $\pm$ 0.050}\\
				\hline
				\textbf{RankSVM-Struct} & \textbf{0.464 $\pm$ 0.022} & 0.470 $\pm$ 0.052\\
				\hline
				\textbf{Random Forest} & 0.458 $\pm$ 0.017 & \textbf{0.473 $\pm$ 0.038}\\
				\hline
			\end{tabular}}
		\end{center}
	\end{table}

		Turning our attention to the  LETOR 4.0 benchmark, Table~\ref{B_COMP_L4_MAP} shows that there are four statistically tied best ranker methods  in the MQ2007 dataset, namely AdaRank-NDCG, ListNet, RankBoost and RankSVM-Struct,  whereas AdaRank-MAP and Random Forest are clear losers. A similar scenario is found in the MQ2008 dataset, but this time RankSVM-Struct is the worst performer and the only one not statistically tied with the others.

		In general, we find that, in the vast majority of the analyzed datasets, most of the baselines are statistically tied, with no clear winner, raising a question of whether it is cost-effective to invest on developing new learning-to-rank algorithms, as opposed to combining multiple methods into a single hybrid solution or investing on reducing the costs (particularly in cases where the L2R methods outperform the single best feature).

	\subsection{Baseline Comparisons - NDCG10}
		\label{sec:COMP_NDCG@10}
		Tables~\ref{B_COMP_L3_NDCG@10} and \ref{B_COMP_L4_NDCG@10} show the comparison among the various considered baselines in LETOR 3.0 and LETOR 4.0, respectively, in terms of average NDCG@10. Once again, results in bold  indicate  the best rankers for each dataset. These results
contribute to strenghthen the conclusions drawn in Section \ref{sec:COMP_MAP}. 
 In particular, we find that almost all algorithms are statistically tied as the best rankers (77\%) in all datasets. Even in the TD2004 dataset, where the best rankers in terms of MAP are FRank, RankBoost and Random Forest, in terms of average NDCG@10, around half of the reported baselines are statistically tied as best solutions. Similarly, nearly all algorithms in TD2003 and NP2003, aside from RankSVM and Regression, respectively, are tied as best rankers, further raising the question of whether or not there  exists an undisputed winner, both by absolute values and statistical significance.

	\begin{table*}[!htb]
		\begin{center}
			\centering \scriptsize
			\caption{Baselines' average NDCG@10 and confidence intervals across the different datasets, LETOR 3.0}
			\label{B_COMP_L3_NDCG@10}
			\centerline{\begin{tabular}{|p{1.4cm}|c|c|c|c|c|c|}
				\hline
				\textbf{Dataset} &\textbf{HP2003} &\textbf{HP2004} &\textbf{NP2003} &\textbf{NP2004} &\textbf{TD2003} &\textbf{TD2004} \\
				\hline
				\textbf{AdaRank-MAP} & \textbf{0.838 $\pm$ 0.065} & \textbf{0.833 $\pm$ 0.065} & \textbf{0.764 $\pm$ 0.073} & \textbf{0.750 $\pm$ 0.061} & \textbf{0.307 $\pm$ 0.129} & \textbf{0.328 $\pm$ 0.078} \\
				\hline
				\textbf{AdaRank-NDCG} & \textbf{0.806 $\pm$ 0.116} & \textbf{0.806 $\pm$ 0.047} & \textbf{0.767 $\pm$ 0.061} & 0.738 $\pm$ 0.051 & \textbf{0.304 $\pm$ 0.156} & \textbf{0.316 $\pm$ 0.071} \\
				\hline
				\textbf{FRank} & 0.797 $\pm$ 0.063 & 0.762 $\pm$ 0.087 & \textbf{0.776 $\pm$ 0.055} & \textbf{0.730 $\pm$ 0.106} & \textbf{0.269 $\pm$ 0.109} & \textbf{0.333 $\pm$ 0.065} \\
				\hline
				\textbf{ListNet} & \textbf{0.837 $\pm$ 0.073} & \textbf{0.784 $\pm$ 0.100} & \textbf{0.802 $\pm$ 0.065} & \textbf{0.813 $\pm$ 0.100} & \textbf{0.348 $\pm$ 0.112} & \textbf{0.318 $\pm$ 0.011} \\
				\hline
				\textbf{RankBoost} & \textbf{0.817 $\pm$ 0.084} & \textbf{0.743 $\pm$ 0.055} & \textbf{0.807 $\pm$ 0.036} & 0.691 $\pm$ 0.101 & \textbf{0.312 $\pm$ 0.120} & \textbf{0.350 $\pm$ 0.043} \\
				\hline
				\textbf{RankSVM} & 0.808 $\pm$ 0.068 & \textbf{0.769 $\pm$ 0.098} & \textbf{0.800 $\pm$ 0.073} & \textbf{0.806 $\pm$ 0.124} & \textbf{0.346 $\pm$ 0.133} & 0.308 $\pm$ 0.024 \\
				\hline
				\textbf{RankSVM-Primal} & \textbf{0.818 $\pm$ 0.082} & \textbf{0.772 $\pm$ 0.091} & \textbf{0.789 $\pm$ 0.059} & \textbf{0.795 $\pm$ 0.110} & \textbf{0.357 $\pm$ 0.127} & 0.291 $\pm$ 0.033 \\
				\hline
				\textbf{RankSVM-Struct} & \textbf{0.816 $\pm$ 0.094} & \textbf{0.767 $\pm$ 0.102} & \textbf{0.795 $\pm$ 0.044} & \textbf{0.798 $\pm$ 0.108} & \textbf{0.347 $\pm$ 0.139} & 0.309 $\pm$ 0.028 \\
				\hline
				\textbf{Regression} & 0.594 $\pm$ 0.061 & 0.647 $\pm$ 0.129 & 0.666 $\pm$ 0.092 & 0.654 $\pm$ 0.149 & \textbf{0.326 $\pm$ 0.107} & 0.303 $\pm$ 0.037 \\
				\hline
				\textbf{Regression-L2reg} & \textbf{0.822 $\pm$ 0.119} & 0.719 $\pm$ 0.071 & \textbf{0.802 $\pm$ 0.046} & \textbf{0.804 $\pm$ 0.106} & \textbf{0.330 $\pm$ 0.144} & 0.283 $\pm$ 0.031 \\
				\hline
				\textbf{Smooth Rank} & \textbf{0.799 $\pm$ 0.104} & \textbf{0.806 $\pm$ 0.098} & \textbf{0.797 $\pm$ 0.052} & \textbf{0.808 $\pm$ 0.137} & \textbf{0.328 $\pm$ 0.127} & 0.291 $\pm$ 0.028 \\
				\hline
				\textbf{SVMMAP} & \textbf{0.832 $\pm$ 0.072} & \textbf{0.822 $\pm$ 0.053} & \textbf{0.799 $\pm$ 0.046} & \textbf{0.807 $\pm$ 0.092} & \textbf{0.337 $\pm$ 0.113} & \textbf{0.334 $\pm$ 0.032} \\
				\hline
				\textbf{Random Forest}& \textbf{0.830 $\pm$ 0.077}& 0.716 $\pm$ 0.095 & \textbf{0.795 $\pm$ 0.040} & 0.708 $\pm$ 0.135& \textbf{0.354 $\pm$ 0.098} & \textbf{0.350 $\pm$ 0.048} \\
				\hline
			\end{tabular}}
		\end{center}
	\end{table*}

	\begin{table}[!hb]
		\begin{center}
			\centering \scriptsize
			\caption{Baselines' average NDCG@10 and confidence intervals across the different datasets, LETOR 4.0}
			\label{B_COMP_L4_NDCG@10}
			\centerline{\begin{tabular}{|c|c|c|}
				\hline
				\textbf{Dataset} & \textbf{MQ2007} & \textbf{MQ2008}\\
				\hline
				\textbf{AdaRank-MAP} & 0.433 $\pm$ 0.028 & \textbf{0.229 $\pm$ 0.055}\\
				\hline
				\textbf{AdaRank-NDCG} & 0.437 $\pm$ 0.029 & \textbf{0.231 $\pm$ 0.055}\\
				\hline
				\textbf{ListNet} & \textbf{0.444 $\pm$ 0.027} & \textbf{0.230 $\pm$ 0.056}\\
				\hline
				\textbf{RankBoost} & \textbf{0.446 $\pm$ 0.029} & \textbf{0.226 $\pm$ 0.053}\\
				\hline
				\textbf{RankSVM-Struct} & \textbf{0.444 $\pm$ 0.031} & \textbf{0.228 $\pm$ 0.054}\\
				\hline
				\textbf{Random Forest} & 0.440 $\pm$ 0.024 & \textbf{0.228 $\pm$ 0.048}\\
				\hline
			\end{tabular}}
		\end{center}
	\end{table}

	Finally, it is worth noting that  Random Forest has a very similar behavior to the one reported in the previous section. Indeed, considering all MAP and NDCG results, Random Forest is tied with the best ranker in 10 out of all 16 evaluated results,  showing the potential of this method.

\section{Conclusions and Future Work}
	\label{sec:Conclusions}
	After almost a decade of research and development of L2R algorithms, we have here raised two controversial but important questions that should be further discussed by the Information Retrieval community.
	First, given all the costs involved in L2R (e.g., labeling, training, tuning) and the overhead introduced by applying such techniques, for instance, for re-ranking search top results at query time, the cost-benefit ratio of applying such algorithms should be further investigated.
	Secondly, given the similar performance, with 95\% confidence,  of most of the 13 selected  L2R algorithms across various different datasets, researching and developing new L2R algorithms may not be worth the effort. Indeed, although in a few datasets there are undisputed best rankers, it is not the case in most analyzed datasets, in neither metric considered here (i.e., MAP and average NDCG@10).

	Rather than providing definitive answers, our goal here is to instigate discussion and re-evaluation of many L2R algorithms after having applied solid statistical methods in our own investigations of the subject.
	As future work, we intend to expand this study to consider even larger datasets, such as the ones provided by Microsoft Learning to Rank and Yahoo! Labs, as well as new L2R algorithms and ranking metrics.

\bibliographystyle{abbrv}
\bibliography{jidmb}

\end{document}